\documentclass[11pt,a4paper]{article}
\RequirePackage[english]{babel}
\RequirePackage[T1]{fontenc}
\RequirePackage[dvips]{graphicx}
\begin{document}

\title{\begin{flushright}
Nikhef 2011-019  
\end{flushright} 
\vspace{10mm}
Intermediate models for longitudinal profiles of cosmic showers}
\author{J.M.C. Montanus \\
\\ \\Nikhef}
\date{dec 7, 2011}
\maketitle

\begin{abstract}
Cosmic rays impacting on the atmosphere cause particle-showers. Several descriptions exist for the evolution of the shower size along the atmospheric depth. The well known functions for shower profiles, Greisen, Gaisser-Hillas and `Gaussian in Age', are intimately connected in that they all are approximate solutions of versions of the Rossi and Greisen diffusion equations. The mathematical connection will be demonstrated by means of two simple models for the longitudinal electromagnetic shower profile. Both models can be regarded either as a generalization of the Heitler model or as a simplification of the diffusion model of Rossi and Greisen. These models are far closer to reality than the Heitler model, while they are not as close to reality as the model of Rossi and Greisen. Therefore, they will be referred to as intermediate models. For each intermediate model the evolution of the shower is governed by either a single differential equation or a single integro-differential equation. The approximate solution of the differential equation is a Gaisser-Hillas function and can be adjusted such that it almost matches the Greisen profile. The approximate solution of the integro-differential equation is a `Gaussian in Age' function. The corresponding profile is, after suitable adjustment, in excellent agreement with the Greisen profile. The analysis also leads to an alternative functional form for the age parameter.

\end{abstract}

\newpage
\section{Introduction}
The longitudinal development of electromagnetic showers can be described by a system of diffusion equations. They can be solved by means of functional transforms and a second order saddlepoint approximation \cite{{1},{2},{3}}. The solution of Rossi and Greisen can be elaborated to what is known as the Greisen function \cite{4}. For this some further approximations had to be made. Because of the inaccuracies involved in these approximations one may ask if a satisfying trial function can also be obtained from a less accurate approach \cite{5}. This is valid as long as the inaccuracy in the model does not disturb too much the essential shape of the profile. Fortunately, as we will see, it does not. Deviations in height and width can be easily adjusted for. In this way we obtain a simple route to the construction of trial functions for shower profiles.
\\
In this paper we will restrict ourselves to the longitudinal development of the electromagnetic cascade governed by the three elementary processes of pair production, Bremsstrahlung and ionization losses. The corresponding shower consists of three particles: electrons, positrons and photons. A simple model for the longitudinal development of the electromagnetic cascade is the Heitler model \cite{6}. More recently, the Heitler model has been applied for the development of the hadronic portion in the initial stages of extensive air showers \cite{7}. According to the Heitler model each particle will split after travelling the same distance into two particles of half the energy of the parent particle. This distance is the splitting length $d=\lambda_r\ln2$, where $\lambda_r$ is the radiation length in the atmosphere: $\lambda_r \approx 37.1 \textrm{ g/cm}^2$ and $d\approx 25.7 \textrm{ g/cm}^2$. We start at atmospheric depth $X=0$. After the first collision there will be 2 particles at atmosferic depth $X=d$. The subsequent collisions lead to 4 particles at $X=2d$, 8 particles at $X=3d$ and so on. At atmospheric depth $nd$ there will be $2^n$ particles in the shower. At this depth the energy of each particle is $E_0 /2^n$, where $E_0$ is the energy of the primary particle. This cascade continues until the energy of the particles falls below the critical value $E_c=84$ MeV, the energy at which the ionization loss is equal to the collisional energy loss. The shower then stops, according to the Heitler model, when $n> n_c$, where $n_c=\ln\left(E_0/E_c\right)/\ln2$.

\section{An intermediate shower model}
In reality particles do not travel equal distances before they split. To model this we cut the atmosphere into slices $\Delta X$ of equal atmospheric thickness. Travelling through a slice,  a \emph{step} from now on, each particle with a certain energy has a chance $p$ to split into 2 particles of half that energy, and a chance $q=1-p$ to continue as a single particle with the original energy. This is a generalization of the Heitler model. After $n$ steps the energies $E(k)$ the particles can posses are
\begin{equation}\label{1}
E(k)=\frac{1}{2^k}E_0,
\end{equation}
where only the levels with $0\leq k\leq n$ can be occupied.
After $n$ steps the expected number of particles in level $k$ is $N(k,n)$. The conservation of energy then requires
\begin{equation}\label{2}
\sum_{k=0}^n \frac{1}{2^k}N(k,n)=1.
\end{equation}
Since a fraction $q$ remains in the energy level and a fraction $2pN$ enter from the higher energy level the expectation values $N(k,n+1)$ are related to the $N(k,n)$ as follows
\begin{equation}\label{3}
N(k,n+1)=qN(k,n)+2pN(k-1,n).
\end{equation}
This relation obeys energy conservation:
\begin{equation}\label{4}
\sum_{k=0}^n\frac{1}{2^k}N(k,n)=1\qquad \Rightarrow \qquad \sum_{k=0}^{n+1}\frac{1}{2^k}N(k,n+1)=1.
\end{equation}
As can be verified by induction, the solution of eqs. (\ref{3}) and (\ref{4}) is
\begin{equation}\label{5}
N(k,n)={n\choose k} (2p)^k q^{n-k}.
\end{equation}
This is the same as having $2^k$ particles of energy $E_0/2^k$ with binomial probability
\begin{equation}\label{6}
P(k,n)={n\choose k}p^kq^{n-k}.
\end{equation}
That is, the splittings are binomially distributed, while $k$ splittings lead to $2^k$ particles. The \emph{total} expected number of particles after $n$ steps is 
\begin{equation}\label{7}
N(n)=\sum_{k=0}^n{n\choose k}(2p)^kq^{n-k}=(2p+q)^n = (1+p)^n.
\end{equation}
Thus $N(n)$ initially grows exponentially.
The expected \emph{fraction} of particles at energy level $k$ after $n$ steps then is
\begin{equation}\label{8}
f(k,n)=\frac{N(k,n)}{N(n)}={n\choose k}\frac{(2p)^k(1-p)^{n-k}}{(1+p)^n}.
\end{equation}
This is actually a binomial distribution with redefined probability:
\begin{equation}\label{9}
\overline p \equiv \frac{2p}{1+p}\qquad \Rightarrow \qquad f(k,n)={n\choose k}( \overline p)^k(1-\overline p)^{n-k}.
\end{equation}
As in each splitting the energy is conserved, $E_n =E_0$, the average energy per particle after $n$ steps is
\begin{equation}\label{10}
\epsilon (n) \equiv \frac{E(n)}{N(n)}=\frac{E_0}{(1+p)^n}.
\end{equation}
For the average number of steps it takes  before a particle splits we consider the probability $w_l$ for a particle to survive $l$ steps, but not $l+1$ steps:
\begin{equation}\label{11}
w_l=q^l p.
\end{equation}
These probabilities are also properly normalized:
\begin{equation}\label{12}
\sum_{l=0}^\infty w_l =p \sum_{l=0}^\infty q^l =\frac{p}{1-q}=1.
\end{equation}
For the average number of steps a particle survives without splitting we then obtain
\begin{equation}\label{13}
<l> =\sum_{l=0}^\infty lw_l =pq\frac{\mbox{d}}{\mbox{d}q}\sum_{l=0}^\infty q^l=pq\frac{\mbox{d}}{\mbox{d}q} \frac{1}{1-q}=\frac{pq}{(1-q)^2}=\frac{q}{p}=\frac{1}{p}-1.
\end{equation}
Since $\left( <l> +1 \right)$ times $ \Delta X$ is the splitting length $d$ and since $n$ times $\Delta X$ is the actual atmospheric depth $X$, we obtain for small values of $p$ the following relation:
\begin{equation}\label{14}
d= \left( <l>+1 \right) \Delta X=\frac{\Delta X}{p}=\frac{X}{np}. 
\end{equation}
For $p=1$ the original Heitler model is retained: $X=nd$. Our aim is to consider the situation in the limit $p \rightarrow 0$.

\section{Continuum limit}
In the continuum limit, $p \rightarrow 0$ and $n \rightarrow \infty$ satisfying (\ref{14}), the difference eq. (\ref{3}) turns into a differential equation which can be written either as
\begin{equation}\label{15}
\frac{\partial N(k,n)}{\partial n}=qN(k,n)+2pN(k-1,n)-N(k,n)=2pN(k-1,n)-pN(k,n)
\end{equation}
or as
\begin{equation}\label{16}
\frac{\partial N(k,X)}{\partial X}=\frac{2}{d}N(k-1,X)-\frac{1}{d}N(k,X).
\end{equation}
Since the binomial distribution in the limit $p\rightarrow 0$ equals the Poisson distribution, one can expect solutions of the form $2^k$ times a Poisson distribution. Indeed the equations (\ref{15}) and (\ref{16}) are solved by 
\begin{equation}\label{17}
N(k,n)=\frac{1}{k!}(2np)^ke^{-np}
\end{equation}
and
\begin{equation}\label{18}
N(k,X)=\frac{1}{k!}\left(\frac{2X}{d}\right)^ke^{-\frac{X}{d}}
\end{equation}
respectively. These solutions satisfy the boundary condition $N(0,0)=1$. Notice that the number of particles in energy level $E(k)=E_0/2^k$ reaches its maximum at depth $X_{max}=kd$. Since $d=\lambda_r \ln 2$ this is $X_{max}=\lambda_r \ln (E_0 /E)$.
The expression for $N(k,n)$ or $N(k,X)$ is the same as having $2^k$ particles of energy $E_0/2^k$ with Poisson distributed probability:
\begin{equation}\label{19}
P(k,n)=\frac{1}{k!}(np)^ke^{-np},
\end{equation}
or
\begin{equation}\label{20}
P(k,X)=\frac{1}{k!}\left(\frac{X}{d}\right)^ke^{-\frac{X}{d}}.
\end{equation}
In other words, the splittings are Poisson distributed with average length $d$, while $k$ splittings lead to $2^k$ particles. We will refer to (\ref{18}) as the Poisson-related distribution.\\
Without absorption the \emph{total} expected number of particles at atmospheric depth $X$, thus after $n$ steps, is 
\begin{equation}\label{21}
N(X)=\sum_{k=0}^n \frac{1}{k!}\left(\frac{2X}{d}\right)^ke^{-\frac{X}{d}},
\end{equation}
which in practice becomes
\begin{equation}\label{22}
N(X)\approx \sum_{k=0}^{\infty} \frac{1}{k!}\left(\frac{2X}{d}\right)^ke^{-\frac{X}{d}}=e^{\frac{X}{d}}.
\end{equation}
That is, without absorption the total number of particles grows as $e^{X/d}$. 

\section{Absorption}
The shower model presented above is applicable to the early stages of the shower, when all particles have relatively large energy. At later stages a particle will be absorbed or scattered out of the shower when its energy drops below the critical value $E_c \simeq 84$ MeV. Then the energy of the shower is no longer conserved. The net growth of the number of particles will slow down. After reaching a maximum the number of particles will decrease and finally the shower will fade out as far as the particles have not reached the surface of the earth yet. Since the energy of a particle after $k$ splittings is $E_0\cdot 2^{-k}$, the particle is taken out of the shower when $k>\lceil n_c \rceil$, where the critical parameter $n_c$ is given by
\begin{equation}\label{23}
n_c  = \frac{\ln \left(E_0/E_c \right)}{\ln 2}.
\end{equation}
In the discrete model we take the ceiling $m \equiv \lceil n_c \rceil$ as the stopping value since a particle with energy $E_0=2^{n_c} \cdot E_c$ may split $m$ times before the energy is below $E_c$.
As a consequence the average total number of particles develops in time as
\begin{equation}\label{24}
N(n) = \sum_{k=0}^{n} \theta(m -k) 2^k P(k,n),
\end{equation}
where $\theta(x)$ is the Heaviside step function
\begin{equation}\label{25}
\theta(x)=\left\{ \begin{array}{ll} 1, & \mbox{if } x \geq 0;\\ 0, & \mbox{if } x < 0.\end{array} \right.
\end{equation}
The previous result for $N(n)$ holds as long as $n \leq m$:
\begin{equation}\label{26}
N(n\leq m;X) = \sum_{k=0}^n \frac{1}{k!}\left(\frac{2X}{d}\right)^k e^{-X/d}.
\end{equation}
However, for $n > m$ the summation is limited:
\begin{equation}\label{27}
N(n>m;X) = \sum_{k=0}^m \frac{1}{k!}\left(\frac{2X}{d}\right)^k e^{-X/d}.
\end{equation}
Since $n=\frac{1}{p}\frac{X}{d}$ goes to infinity in the limit $p \rightarrow 0$, the latter equation can in practise be used at all depths:
\begin{equation}\label{27a}
N_m (X) = \sum_{k=0}^m \frac{1}{k!}\left(\frac{2X}{d}\right)^k e^{-X/d}.
\end{equation}
It can also be written as 
\begin{equation}\label{27b}
N_m (X) = \frac{d}{2} \cdot e^{X/d} \cdot \sum_{k=0}^m f(X;k+1,d/2)
\end{equation}
or as 
\begin{equation}\label{27c}
N_m (X) =e^{X/d} \cdot \frac{\Gamma (m+1,2X/d)}{\Gamma (m+1)},
\end{equation}
where $f$ is the Gamma distribution
\begin{equation}\label{27d}
f(X;k,\theta) =\frac{X^{k-1} \cdot e^{-X/\theta}}{\theta^k \cdot \Gamma (k)},
\end{equation}
where $\Gamma(m+1,2X/d)$ is the incomplete Gamma function and where $\Gamma$ is the Gamma function: $\Gamma(m+1)=m!$ if $m$ is an integer.
By taking the derivative of expression (\ref{27a}) with respect to $X$ we obtain
\begin{equation}\label{27e}
\frac{d N_m (X)}{d X}= \frac{2}{d} N_{m-1} (X) -\frac{1}{d} N_m (X)
\end{equation}
as well as
\begin{equation}\label{27f}
\frac{d N_m (X)}{d X}= \frac{1}{d} N_{m} (X) -\frac{2}{d}\frac{1}{m!} \left( \frac{2X}{d} \right)^m e^{-X/d}.
\end{equation}
From eqs. (\ref{16}) and (\ref{27e}) we see that the total number of particles satisfies the same differential equation as the number of particles in each energy level. This suggests to take a single term of the Poisson series as an approximation for the total number of particles. Numerical evaluation of the expression (\ref{27a}) demonstrates that the total number of particles reaches a maximum $N_{\max}$ at atmospheric depth
\begin{equation}\label{27g}
X_{\max} \approx \frac{m^2 d}{m+1} \approx (m-1)d.
\end{equation}
Therefore the forelast term of the series (\ref{27a}) seems to be the best choice:
\begin{equation}\label{27h}
N_m (X) \approx \frac{A}{\Gamma (m)} \left( \frac{2X}{d} \right)^{m-1} e^{-X/d},
\end{equation}
where $A$ is a normalization constant. This approximation actually is a Gamma distribution. It is known that a truncated Poisson series can be accurately approximated by a Gamma distribution. Another motivation for the Gamma distribution lies in the fact that it satisfies the same differential equation as the truncated Poisson series. We will make further remark about our motivation for the Gamma distribution at the end of section 8. 
As desired (\ref{27h}) reaches its maximum at $X_{\max} = (m-1)d$. 
From eqn. (\ref{27f}) we find:
\begin{equation}\label{27i}
N_{\max}= \frac{2}{m \Gamma (m) } \left( \frac{2X_{\max}}{d} \right)^m e^{-X_{\max}/d},
\end{equation}
while for the approximation (\ref{27h}) this is
\begin{equation}\label{27j}
N_{\max}\approx \frac{A}{ \Gamma (m) } \left( \frac{2X_{\max}}{d} \right)^{m-1} e^{-X_{\max}/d}.
\end{equation}
Comparison of the latter two eqs. gives
\begin{equation}\label{27z}
A \approx 4 \left( 1-\frac{1}{m} + O\left(\frac{1}{m^2} \right) \right).
\end{equation}
By means of this expression for $A$ and by means of the Stirling approximation, $\Gamma (m) \approx \sqrt{2 \pi } \sqrt{m-1} \left(\frac{m-1}{e} \right)^{m-1}$, the function (\ref{27h}) takes the form
\begin{equation}\label{27k}
N_m (X) \approx g(m) \sqrt{\frac{2}{\pi}}\frac{2^m}{\sqrt{m-\frac{1}{2}}} \left( \frac{X}{(m-1)d} \right)^{m-1} e^{m-1-X/d},
\end{equation}
where 
\begin{equation}\label{27w}
g(x) \approx 1-\frac{3}{4x} + O\left(\frac{1}{x^2} \right).
\end{equation}
Notice that $g$ is close to unity. As an example $g \approx 0.97$ for $m=25$. As known the agreement between a Gamma distribution and a truncated Poisson series can be further improved by means of a parameter $X_0$ as follows: 
\begin{equation}\label{27l}
N_m (X) \approx \sqrt{\frac{2}{\pi}}\frac{2^m}{\sqrt{m-\frac{1}{2}}} \left( \frac{X-X_0}{(m-1)d-X_0} \right)^{m-1-X_0/d} e^{m-1-X/d}.
\end{equation}
Accurate fits are obtained with $X_0 \approx -d/2$. The larger the negative values for $X_0$, the larger the width of the profile, the \emph{fwhm} for instance. Writing $\sqrt{\frac{2}{\pi}}\frac{2^m}{\sqrt{m-\frac{1}{2}}}$ as $N_{\max}$, $(m-1)d$ as $X_{max}$ and $d$ as $\lambda$, the expression (\ref{27l}) reads
\begin{equation}\label{27m}
N_m (X) \approx N_{\max} \left( \frac{X-X_0}{X_{\max}-X_0} \right)^{\frac{X_{\max}-X_0}{\lambda}} e^{\frac{X_{\max}-X_0}{\lambda}}.
\end{equation}
We clearly recognize it as the Gaisser-Hillas function \cite{8}. For the moment we will restrict ourselves to the case $X_0=-\frac{1}{2}d$. We will adjust the width of the profile afterwards. The approximation of a truncated Poisson series by a Gamma distribution is pure mathematics. The parameter $X_0$ therefore is just a mathematical parameter. This (and the fact that its value is preferably negative) supports the opinion that it should not be given a physical interpretation as being the point of first interaction \cite{{9},{10}}.  

The function $m=\lceil n_c \rceil$ approximately goes as $m \approx n_c+\frac{1}{2}$. The substitution of $m \approx n_c+\frac{1}{2}$ in the expression (\ref{27l}) leads to
\begin{equation}\label{34}
N_m \approx g(n_c) \frac{2}{\sqrt{\pi}}\frac{2^{n_c}}{\sqrt{n_c}} \left( \frac{X+d/2}{n_c d} \right)^{n_c} e^{n_c -1/2-X/d}.
\end{equation}
To visualize its accuracy both the numerical summation (\ref{27a}) and the function (\ref{34}) are plotted in figure 1.  
\begin{figure}[htbp]
\includegraphics{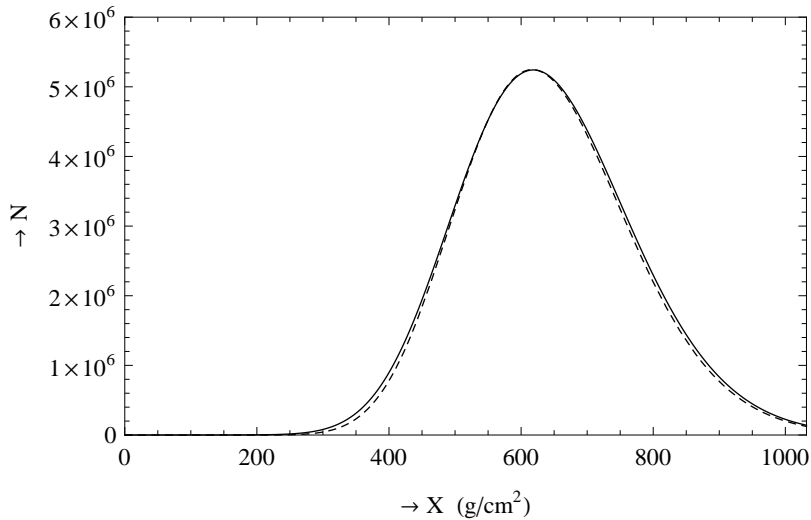}
\caption{Longitudinal shower profiles according to the truncated Poisson series (solid) as well as to its approximation (dashed) for a $2 \cdot 10^{15}$ eV shower. Vertical is the number of particles and horizontal is the atmospheric depth.}
\end{figure}
The expression (\ref{34}) leads to the same profile as the following exression
\begin{equation}\label{35}
N_m (X) \approx \frac{2}{\sqrt{\pi}}\frac{2^{n_c}}{\sqrt{n_c}} \left( \frac{X}{n_c d} \right)^{n_c} e^{n_c-X/d},
\end{equation}
except that it is translated over a distance $d/2$. In the next section we will make a comparison with the Greisen function. Since we are interested in the main characteristics of the shape of the profile and not in the difference caused by a small translation, we will conveniently use expression (\ref{35}) hereafter.

\section{The ratio of the numbers of particles}
Usually one works with an expression for the number of charged particles in the shower instead of the total number of particles. To this end we will distinguish the electrons and positrons from the photons. An electron/positron can split into an electron/positron and a photon, while a photon splits into an electron and a positron. It doesn't change the present model, while we can now look at the following system of difference equations:
\begin{eqnarray}\label{38}
N_{e\pm}(n+1)=N_{e\pm}(n)+2pN_{\gamma}(n), \nonumber \\
N_{\gamma}(n+1)=pN_{e\pm}(n)+(1-p)N_{\gamma}(n).
\end{eqnarray}
Here $N_{e\pm}$ is the number of electrons and positrons and $N_\gamma$ is the number of photons.
For the ratio $\rho=N_{e\pm}/N_{\gamma}$ we find
\begin{eqnarray}\label{39}
\rho(n+1)N_{\gamma}(n+1)=\rho(n)N_{\gamma}(n)+2pN_{\gamma}(n), \nonumber \\
N_{\gamma}(n+1)=p\rho(n)N_{\gamma}(n)+(1-p)N_{\gamma}(n).
\end{eqnarray}
The elimination of $N_\gamma$ leads to
\begin{equation}\label{40}
\rho(n+1)=\frac{\rho(n)+2p}{1-p+p\rho(n)}.
\end{equation}
The ratio $\rho$ asymptotically approaches a limit value $R$, which satisfies $R^2-R-2=0$. The latter equation is solved by the stable stationary point $R=2$. From $N_{e\pm}=2N_{\gamma}$ we obtain $N_\gamma=\frac{1}{3}N$ and $N_{e\pm}=\frac{2}{3}N$. To obtain the number of charged particles we therefore have to multiply the expression (\ref{35}) by $\frac{2}{3}$. So, under the assumption that the shower develops with the ratio in its equilibrium according to the intermediate model the number of charged particles is given by
\begin{equation}\label{41}
N_{e\pm}(X)= g(n_c) \frac{4}{3\sqrt{\pi}} \frac{2^{n_c}}{\sqrt{n_c}} \left( \frac{eX}{n_c d} \right)^{n_c}\cdot e^{-\frac{X}{d}}.
\end{equation}

\section{Expressions for the shower profile}
The shower profile according to the expression (\ref{41}) turns out to be twice too high, twice too narrow compared to, for instance, the Greisen profile. A deviation from the Greisen profile could be expected since in reality there will also be other than fifty-fifty splittings with different probabilities for different ratios. Moreover, for Bremsstrahlung these probabilities differ from the ones for pair creation. If this is taken into account, one arrives at the system of diffusion equations of Rossi and Greisen \cite{1}. We will return to it in the next section. Here we will modify the eq. (\ref{41}) such that the height and width of the corresponding profile is in agreement with the Greisen profile. We halve the height by dividing (\ref{41}) by $2$. We double the width of the profile by multiplying the powers in (\ref{41}) by $\frac{2}{3}\ln 2$. The result is
\begin{equation}\label{43}
N_{e\pm}(X)= g(n_c) \frac{2}{3\sqrt{\pi}} \frac{2^{n_c}}{\sqrt{n_c}} \cdot \left( \frac{eX}{n_c d} \right)^{\frac{2}{3} n_c \ln 2}\cdot e^{-\frac{2X\ln 2}{3d}}.
\end{equation}
This still is a Gaiser-Hillas type of function as can be verified by substituting $X_{max}=n_c d$, $\lambda =\frac{3d}{2\ln 2}$ and $X_0 =0$ into expression (\ref{27m}).
The eqn. (\ref{43}) can also be written as
\begin{equation}\label{44}
N_{e\pm}(t)= g(n_c) \frac{2}{3\sqrt{\pi}} \frac{2^{n_c}}{\sqrt{n_c}} \cdot \left( \frac{et}{n_c \ln 2} \right)^{\frac{2}{3}n_c \ln 2}\cdot e^ {-\frac{2}{3}t},
\end{equation}
where $t=\frac{X}{\lambda_r}$ is the atmospheric depth in units of radiation length. Its maximum value 
\begin{equation}\label{45}
N_{e\pm,\max}=g(n_c) \frac{2}{3\sqrt{\pi}} \frac{2^{n_c}}{\sqrt{n_c}}
\end{equation}
occurs at depth $t_{\max} =n_c \ln 2$.

The well-known Greisen approximation formula reads \cite{2}:
\begin{equation}\label{46}
N_{e\pm}(t)=\frac{0.31}{\sqrt{y_c}}\cdot  e^{t\cdot \left( 1-\frac{3}{2}\ln s \right)},
\end{equation}
where $s=\frac{3t}{t+2y_c}$ is the \emph{age-parameter} and $y_c=ln\left(E_0/E_c\right)\equiv n_c \ln 2$.
Completely in terms of $t$ and $n_c$ this is
\begin{equation}\label{47}
N_{e\pm}(t)=\frac{0.31}{\sqrt{n_c \ln 2}}\cdot \left(\frac{1}{3}+\frac{2}{3}\frac{n_c \ln 2}{t}\right)^{\frac{3}{2}t}\cdot e^t.
\end{equation}
Its maximum value 
\begin{equation}\label{48}
N_{e\pm,\max}=\frac{0.31}{\sqrt{\ln 2}}\cdot \frac{2^{n_c}}{\sqrt{n_c}}\approx 0.37\cdot \frac{2^{n_c}}{\sqrt{n_c}}
\end{equation}
also occurs at depth $t_{\max} =n_c \ln 2$. 
For a visual comparison both the present Gaisser-Hillas profile and the Greisen profile are plotted in the figure 2. It shows that both profiles practically coincide. From the comparison of (\ref{45}) with (\ref{48}) we conclude that the value $0.31$ in the Greisen function approximately equals the semi-theoretical value $\frac{2 \sqrt{ \ln 2}}{3\sqrt{\pi}}$. 
 
\begin{figure}[htbp]
\includegraphics{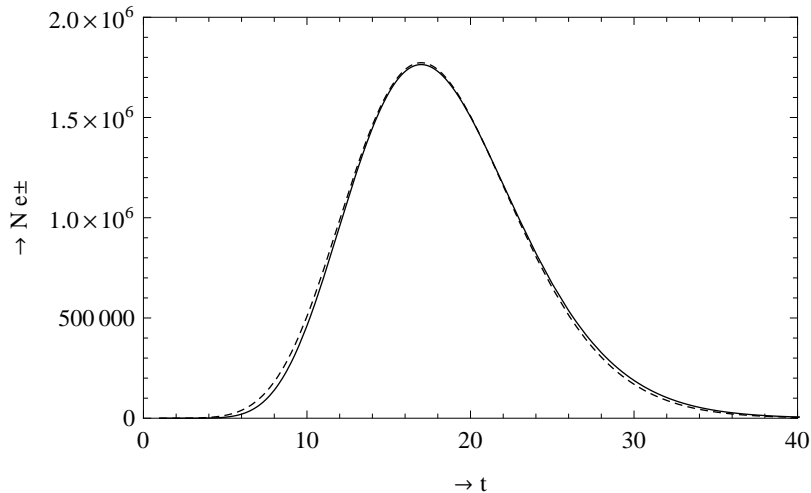}
\caption{Longitudinal shower profiles according to the present Gaisser-Hillas function (solid) and the Greisen function (dashed) for a $2 \cdot 10^{15}$ eV shower. Vertical is the number of electrons and positrons and horizontal the atmospheric depth (in units of radiation length).}
\end{figure}

\section{The connection with Rossi and Greisen}
In this section we will show the connection between the preceeding model and the cosmic-ray theory of Rossi and Greisen \cite{1}. For our purposes it suffices to consider the situation under what is known as the `approximation A'. Under this approximation the complete screening cross sections of radiation and pair creation processes are used. Other processes like Compton scattering are neglected and ionization loss is solely used as a stopping criterion \cite{{1},{3}}. The diffusion equations for the differential distributions $n_{e\pm}$ and $n_{\gamma}$ then read
\begin{eqnarray}\label{49}
\frac{\partial n_{e\pm}(E,t)}{\partial t}=2\int_E^\infty n_{\gamma}(W,t)\frac{1}{W}\psi_0 \left(\frac{E}{W}\right)\mbox{d}W+ \nonumber \\ \int_E^\infty n_{e\pm}(E',t)\frac{1}{E'}\varphi_0 \left(\frac{E'-E}{E'}\right)\mbox{d}E' -\int_0^E n_{e\pm} (E,t)\frac{1}{E}\varphi_0 \left(\frac{W}{E}\right)\mbox{d}W, \qquad
\end{eqnarray}
\begin{equation}\label{50}
\frac{\partial n_{\gamma}(W,t)}{\partial t}=\int_W^\infty n_{e\pm}(E,t)\frac{1}{E}\varphi_0 \left(\frac{W}{E} \right) \mbox{d}E -\int_0^W n_{\gamma}(W,t)\frac{1}{W} \psi_0 \left(\frac{E}{W}\right)\mbox{d}E.
\end{equation}
In these equations $\varphi_0 \left(\frac{W}{E}\right)\frac{1}{E}\mbox{d}W$ is the differential probability per radiation length for an electron (or a positron) with energy between $E$ and $E+\mbox{d}E$ to split of a photon with energy between $W$ and $W+\mbox{d}W$. Similarly, $\psi_0 \left(\frac{E}{W}\right)\frac{1}{W}\mbox{d}E$ is the differential probability per radiation length for a photon with energy between $W$ and $W+\mbox{d}W$ to produce a pair with the electron energy between $E$ and $E+\mbox{d}E$. They are given by \cite{1} :
\begin{equation}\label{51}
\psi_0(u)=u\varphi_0(1/u)=\frac{4}{3}u^2-\frac{4}{3}u+1+2b(u^2-u).
\end{equation}
The last term on the rhs is of minor importance since $b$ is relatively small, $b \approx 0.013$.
After a suitable change of variables the equations (\ref{49}) and (\ref{50}) take the form
\begin{eqnarray}\label{52}
\frac{\partial n_{e\pm}(E,t)}{\partial t}=2\int_0^1 n_{\gamma} \left(\frac{E}{u},t\right)\frac{ \psi_0 (u)}{u}\mbox{d}u+\int_0^1 n_{e\pm} \left(\frac{E}{1-v},t\right)\frac{\varphi_0 (v)}{1-v}\mbox{d}v \nonumber \\ -n_{e\pm}(E,t)\int_0^1\varphi_0 (v)\mbox{d}v, \qquad
\end{eqnarray}
\begin{equation}\label{53}
\frac{\partial n_{\gamma}(W,t)}{\partial t}=\int_0^1 n_{e\pm}\left(\frac{W}{v},t\right)\frac{\varphi_0 (v)}{v}\mbox{d}v -n_{\gamma}(W,t)\int_0^1\psi_0 (u)\mbox{d}u.
\end{equation}
By means of separation of the variables energy and depth one obtains the `elementary' solutions:
$n_{e\pm}(E,t)=aE^{-(s+1)}e^{\lambda t}$ and $n_{\gamma}(W,t)=bW^{-(s+1)}e^{\lambda t}$. Upon substitution a quadratic equation for $\lambda$ is obtained. One solution, $\lambda_2 (s)$, corresponds to a quick adaptation to the equilibrium ratio between $n_{e\pm}$ and $n_{\gamma}$. The other solution, $\lambda_1 (s)$, describes the main development of the shower, see section 27 of reference \cite{1}.
The connection of the intermediate model with the model of Rossi and Greisen follows when delta-functions are substituted for the probabilities. The eqns (\ref{49}) and (\ref{50}) then take the form  
\begin{eqnarray}\label{54}
\frac{\partial n_{e\pm}(E,t)}{\partial t}=2\int_E^\infty n_{\gamma}\left(W,t\right)a\delta (W-2E)\mbox{d}W+\nonumber \\ \int_E^\infty n_{e\pm}\left(E',t\right)a\delta (E'-2E)\mbox{d}E'  -n_{e\pm}(E,t)\int_0^E a\delta (2W-E)\mbox{d}W, \qquad
\end{eqnarray}
\begin{equation}\label{55}
\frac{\partial n_{\gamma}(W,t)}{\partial t}=\int_W^\infty n_{e\pm}\left(E,t\right)a\delta (E-2W)\mbox{d}E -n_{\gamma}(W,t)\int_0^W a\delta (2E-W)\mbox{d}E,
\end{equation}
where $a=\frac{1}{\ln 2}$ since the differential probabilities in the diffusion equations are per radiation length $\lambda_r$, while the splitting probability is a delta-function per splitting distance $d=\lambda_r \ln 2$. The system (\ref{54}) and (\ref{55}) reduces to
\begin{equation}\label{56}
\frac{\partial n_{e\pm}(E,t)}{\partial t}=2a n_\gamma (2E,t)+a n_{e\pm} (2E,t)-a n_{e\pm} (E,t)
\end{equation}
\begin{equation}\label{57}
\frac{\partial n_{\gamma}(W,t)}{\partial t}=a n_{e\pm} (2W,t)-a n_{\gamma}(W,t).
\end{equation}
In the hypothetical one-particle model there also is no difference between photons, electrons or positrons. Given the $1:2$ ratio for the number of photons and electrons/positrons in our model, we take $n_{e\pm}(E,t)=\frac{2}{3}n(E,t)$ and $n_{\gamma}(W,t)=\frac{1}{3}n(E,t)$. We can also set $W$ equal to $E$. Then the system of equations further reduce to
\begin{equation}\label{58}
\frac{2}{3}\frac{\partial n(E,t)}{\partial t}=\frac{2}{3} an(2E,t)+\frac{2}{3} an(2E,t)-\frac{2}{3} an(E,t)
\end{equation}
\begin{equation}\label{59}
\frac{1}{3}\frac{\partial n(E,t)}{\partial t}=\frac{2}{3} a n(2E,t)-\frac{1}{3}an(E,t).
\end{equation} 
Obviously, these equations are identical. As it should, the one-particle model is governed by a single differential equation:
\begin{equation}\label{60}
\frac{\partial n(E,t)}{\partial t}=2 a n(2E,t)-a n(E,t).
\end{equation} 
Substituting $a=\frac{1}{\ln 2}$ we get
\begin{equation}\label{61}
\frac{\partial n(E,t)}{\partial t}=\frac{2}{\ln 2}n(2E,t)-\frac{1}{\ln 2}n(E,t),
\end{equation}
or
\begin{equation}\label{62}
\frac{\partial n(E,X)}{\partial X}=\frac{2}{d}n(2E,X)-\frac{1}{d}n(E,X).
\end{equation}
From $E(k)=E_0 \cdot 2^{-k}$ it follows that $E(k-1)=2E(k)$. Hence,
\begin{equation}\label{63}
\frac{\partial n(E(k),X)}{\partial X}=\frac{2}{d}n(E(k-1),X)-\frac{1}{d}n(E(k),X).
\end{equation}
This is the same differential equation as in (\ref{16}).

\section{Another intermediate model}
In the previous model we considered splittings with the energy equally divided between the decay particles. We obtain a little more accurate model by allowing other than fifty-fifty splittings, although all with equal probability. That is, we take the differential probabilities for the different splittings equal to a constant. It turns out that the maximum shower size occurs at the desired depth if the constant is taken equal to 2. So, for the following analysis we will restrict ourselves to the case $\psi_0(u)=\varphi_0(v)=2$. Then the system of equations (\ref{52}) and (\ref{53}) take the form
\begin{eqnarray}\label{107}
\frac{\partial n_{e\pm}(E,t)}{\partial t}=2\int_0^1 n_{\gamma} \left(\frac{E}{u},t\right)\frac{2}{u}\mbox{d}u+\int_0^1 n_{e\pm} \left(\frac{E}{1-v},t\right)\frac{2}{1-v}\mbox{d}v \nonumber \\ -n_{e\pm}(E,t)\int_0^1 2 \mbox{d}v, \qquad
\end{eqnarray}
\begin{equation}\label{108}
\frac{\partial n_{\gamma}(W,t)}{\partial t}=\int_0^1 n_{e\pm}\left(\frac{W}{v},t\right)\frac{2}{v}\mbox{d}v -n_{\gamma}(W,t)\int_0^1 2 \mbox{d}u.
\end{equation}
After a suitable change of variables this is:
\begin{equation}\label{109}
\frac{\partial n_{e\pm}(E,t)}{\partial t}=2\int_0^1 n_{\gamma} \left(\frac{E}{u},t\right)\frac{2}{u}\mbox{d}u+\int_0^1 n_{e\pm} \left(\frac{E}{u},t\right)\frac{2}{u}\mbox{d}u -2 n_{e\pm}(E,t),
\end{equation}
\begin{equation}\label{110}
\frac{\partial n_{\gamma}(W,t)}{\partial t}=\int_0^1 n_{e\pm}\left(\frac{W}{u},t\right)\frac{2}{u}\mbox{d}u -2 n_{\gamma}(W,t).
\end{equation}
Also in this model we do not distinguish between photons and charged particles. By setting $n_{e\pm}(E,t)=2n_{\gamma}(W,t)=\frac{2}{3}n(E,t)$ and $W=E$, we obtain
\begin{equation}\label{111}
\frac{2}{3}\frac{\partial n(E,t)}{\partial t}=\frac{4}{3} \int_0^1 n \left(\frac{E}{u},t\right)\frac{1}{u}\mbox{d} u+\frac{4}{3} \int_0^1 n\left(\frac{E}{u},t\right)\frac{1}{u}-\frac{4}{3} n(E,t)
\end{equation}
\begin{equation}\label{112}
\frac{1}{3}\frac{\partial n(E,t)}{\partial t}=\frac{4}{3} \int_0^1 n \left(\frac{E}{u},t\right)\frac{1}{u} \mbox{d} u-\frac{2}{3} n(E,t).
\end{equation} 
Obviously, these equations are identical. Also this one-particle model is governed by a single diffusion equation:
\begin{equation}\label{113}
\frac{\partial n(E,t)}{\partial t}=4 \int_0^1 n\left(\frac{E}{u},t\right)\frac{1}{u} \mbox{d} u-2 n(E,t).
\end{equation} 
This equation allows for elementary solutions in which the variables $E$ and $t$ are separated. To be specific, solutions of the type
\begin{equation}\label{114}
n(E,t)=A\cdot \left(\frac{E_0}{E}\right)^{s+1}\cdot e^{\lambda t},
\end{equation}
with $s \neq -1$, do satisfy the differential equation (\ref{113}) if
\begin{equation}\label{115}
\lambda(s) =\frac{4}{s+1}-2.
\end{equation}
Notice that if $s$ does depend on $t$, we would have the additional requirement $y+\lambda't=0$, where $y=\ln (E_0 /E)$ and where the prime stands for the derivation with respect to $s$. This requirement is precisely the saddle point condition (\ref{124}) as we will see soon.
Now we will construct the solution in an analogous manner as in the paper of Rossi and Greisen. To this end we consider the Mellin integral
\begin{equation}\label{116}
M_n(s,t)=\int_0^\infty E^{s} n(E,t)\mbox{d} E
\end{equation}
and its inverse transformation
\begin{equation}\label{117}
n(E,t)=\frac{1}{2\pi i}\int_{c-i\infty}^{c+i\infty} E^{-s-1} M_n(s,t)\mbox{d} s.
\end{equation}
Multiplying both sides of eqn. (\ref{113}) by $E^{s}$ and integrating with respect to energy from $0$ to $\infty$, we obtain: 
\begin{eqnarray}\label{118}
\int_0^\infty E^{s} \frac{\partial n(E,t)}{\partial t}\mbox{d} E=4 \int_0^\infty E^{s} \int_0^1 n\left(\frac{E}{u},t\right)\frac{1}{u}\mbox{d} u \mbox{d} E- \nonumber \\ 2 \int_0^\infty E^{s} n(E,t)\mbox{d} E.
\end{eqnarray}
Since $n(E/u,t)=u^{(s+1)} n(E,t)$, the latter is reduced to
\begin{eqnarray}\label{119}
\frac{\partial}{\partial t}\int_0^\infty E^{s} n(E,t) \mbox{d} E=\frac{4}{s+1} \int_0^\infty (E)^{s} n(E,t)\mbox{d} (E)- \nonumber \\ 2 \int_0^\infty E^{s} n(E,t)\mbox{d} E.
\end{eqnarray}
Hence,
\begin{equation}\label{120}
\frac{\partial}{\partial t}M_n(s,t)=\lambda(s) M_n(s,t),
\end{equation}
with $\lambda(s)$ as given by (\ref{115}).
The solution of this equation is
\begin{equation}\label{121}
M_n(s,t)=M_n(s,0)\cdot e^{\lambda (s) t},
\end{equation}
where 
\begin{equation}\label{122}
M_n(s,0)=\int_0^\infty E^{s} n(E,0)\mbox{d}E=\int_0^\infty E^{s} \delta (E-E_0)\mbox{d}E=E_0^{s},
\end{equation}
Next we apply the inverse Mellin transformation:
\begin{eqnarray}\label{123}
n(E,t)=\frac{1}{2\pi i}\int_{c-i\infty}^{c+i\infty}E^{-s-1}M_n(s,t)\mbox{d}s \nonumber \\
=\frac{1}{2\pi i}\int_{c-i\infty}^{c+i\infty}E^{-s-1} E_0^{s} e^{\lambda(s)t}\mbox{d}s \nonumber \\
=\frac{1}{2\pi i} \frac{1}{E_0} \int_{c-i\infty}^{c+i\infty}\left(\frac{E_0}{E}\right)^{s+1} e^{\lambda(s)t}\mbox{d}s \nonumber \\
=\frac{1}{E_0}\frac{1}{2 \pi i}\int_{c-i\infty}^{c+i\infty} e^{y\cdot (s+1)+\lambda(s)t}\mbox{d}s,
\end{eqnarray}
where $y=\ln (E_0/E)=k \ln 2$ is the \emph{lethargy}. The real constant $c$ must be in the strip of analyticity, which is the positive halfplane.
We already saw in section 3 that the distribution in energy level $E$ reaches a maximum at $t_{max}=y$. At this depth the exponent $y \cdot (s+1)+\lambda(s)t$ is equal to $t_{max}\cdot (s+1)+\lambda(s)t_{max}$. For $s$ along the real axis it has a minimum at $\bar{s}$ given by $\lambda'(\bar{s})=-1$, That is, for $\bar{s}=1$ and thus $\lambda =0$. Since an analytic function satisfies the Cauchy-Riemann equations and thus the Laplace equation, the exponential term should have a maximum at the point $\bar{s}$, along directions perpendicular to the real axis. Although one usually does not know in advance the relation between $y$ and $t_{max}$, the foregoing makes it clear that we can require the integrand to have a saddle point at the point $\bar{s}$ defined by
\begin{equation}\label{124}
y+\lambda'(\bar{s})t=0,
\end{equation}
where the prime stands for differentiation with respect to $s$. A second order Taylor series of the exponent of the integrand around the point $\bar{s}$ then yields
\begin{equation}\label{125}
y\cdot(s+1)+\lambda(s)t \approx y\bar{s}+y+\lambda(\bar{s})t+\frac{1}{2}\lambda''(\bar{s})t(s-\bar{s})^2.
\end{equation}
Taking the integration path through the saddlepoint, we obtain
\begin{equation}\label{126}
n(E,t)=\frac{1}{E_0}\frac{1}{2 \pi i}\int_{\bar{s}-i\infty}^{\bar{s}+i\infty} e^{y\bar{s}+y+\lambda(\bar{s})t+\frac{1}{2}\lambda''(\bar{s})t(s-\bar{s})^2}\mbox{d}s.
\end{equation}
With the change of variables, $s=\bar{s}+ix$, this is
\begin{equation}\label{127}
n(E,t)=\frac{1}{E_0}\frac{1}{2 \pi}e^{y\bar{s}+y+\lambda(\bar{s})t}\int_{-\infty}^{\infty} e^{-\frac{1}{2}\lambda''(\bar{s})tx^2}\mbox{d}x.
\end{equation}
Evaluating the Gaussian integral, we obtain
\begin{equation}\label{128}
n(E,t)=\frac{1}{E_0} \frac{e^{y\bar{s}+y+\lambda(\bar{s})t}}{\sqrt{2\pi \lambda''(\bar{s})t}}.
\end{equation}
Next we require the solutions to reach a maximum. The function $e^{y\bar{s}+y+\lambda(\bar{s})t}$ reaches its maximum at depth $t_{max}$ given by
\begin{equation}\label{129}
\left[y+\lambda '(\bar{s})t_{max}\right]\left(\frac{\mbox{d}\bar{s}}{\mbox{d}t}\right)_{t=t_{max}}+\lambda(\bar{s})=0.
\end{equation}
Because of the saddlepoint relation (\ref{124}) the latter implies $\lambda(\bar{s})=0$. From (\ref{115}) it is inferred that the maximum occurs at $\bar{s}=1$. Since $\lambda'(1)=-1$ it follows from (\ref{124}) that $t_{max}=y$ or $X_{max}=\lambda_r \ln (E_0/E)$ as desired.
From (\ref{124}) and (\ref{115}) it follows that 
\begin{equation}\label{130}
\lambda''(\bar{s})t=y\sqrt{\frac{y}{t}},
\end{equation}
\begin{equation}\label{131}
\lambda(\bar{s})t=2\sqrt{ty}-2t
\end{equation}
and
\begin{equation}\label{132}
\bar{s}=2\sqrt{\frac{t}{y}}-1.
\end{equation}
Hence,
\begin{equation}\label{133}
n(E,t)=\frac{1}{E_0} \frac{e^{4\sqrt{ty}-2t}}{\sqrt{2\pi \cdot y}}.
\end{equation}
For the integral distribution we can also make use of the Mellin transform:
\begin{eqnarray}\label{134}
M_N (s-1,t)=\int_0^\infty E^{s-1}N(W>E,t)\mbox{d}E \leftrightarrow \nonumber \\ 
N(W>E,t)=\frac{1}{2 \pi i}\int_{c-i\infty}^{c+i\infty} E^{-s} M_N (s-1,t)\mbox{d}s.
\end{eqnarray}
As can be verified by means of partial integration and the property $\frac{\mbox{d}}{\mbox{d}y}\int_y^\infty f(x)\mbox{d}x=-f(y)$, there holds the following relation between the Mellin transforms of the integral and differential distribution:
\begin{equation}\label{135}
M_N (s-1,t)=\frac{1}{s} M_n (s,t),
\end{equation}
where $N=N (W>E,t)= \int_E^\infty  n(E',t)\mbox{d}E'$. From this relation we obtain
\begin{equation}\label{136}
N(W>E,t)=\frac{1}{2 \pi i}\int_{c-i\infty}^{c+i\infty} \frac{1}{s} E^{-s} M_n (s,t)\mbox{d}s.
\end{equation}
Substitution of the expression (\ref{121}) for $M_n$ leads to
\begin{equation}\label{137}
\frac{1}{2 \pi i}\int_{c-i\infty}^{c+i\infty} \frac{1}{s} \left(\frac{E_0}{E}\right)^{s} e^{\lambda (s)t} \mbox{d}s=\frac{1}{2 \pi i}\int_{c-i\infty}^{c+i\infty} e^{-\ln s +y\cdot s+\lambda (s)t} \mbox{d}s,
\end{equation}
where again
\begin{equation}\label{138}
\lambda (s)=\frac{4}{s+1}-2.
\end{equation}
From here we can proceed in a similar manner as for the differential distribution. The integrand has a saddle point at the point $\bar{s}$ defined by
\begin{equation}\label{139}
\frac{-1}{\bar{s}}+y+\lambda'(\bar{s})t=0.
\end{equation}
A second order Taylor series of the exponent of the integrand around the point $\bar{s}$ then yields
\begin{equation}\label{140}
-\ln s +y\cdot s+\lambda(s)t \approx -\ln \bar{s} +y\cdot\bar{s}+\lambda(\bar{s})t+\frac{1}{2}r(\bar{s},t)(s-\bar{s})^2,
\end{equation}
where
\begin{equation}\label{141}
r(\bar{s},t)=\left(\frac{1}{\bar{s}^2}+\lambda''(\bar{s})t\right).
\end{equation}
Taking the integration path through the saddlepoint, we obtain to second order
\begin{equation}\label{142}
N(W>E,t)= \frac{1}{2 \pi i}\int_{\bar{s}-i\infty}^{\bar{s}+i\infty} e^{-\ln \bar{s}+y\cdot \bar{s}+\lambda(\bar{s})t+r(\bar{s},t)\frac{1}{2}(s-\bar{s})^2}\mbox{d}s.
\end{equation}
With the change of variables, $s=\bar{s}+ix$, this is
\begin{equation}\label{143}
N(W>E,t)=\frac{1}{2 \pi}e^{-\ln \bar{s}+y\cdot \bar{s}+\lambda(\bar{s})t}\int_{-\infty}^{\infty} e^{-\frac{1}{2}r(\bar{s},t)x^2}\mbox{d}x.
\end{equation}
Evaluating the integral, we obtain
\begin{equation}\label{144}
N(W>E,t)= \frac{e^{-\ln \bar{s}+y\cdot\bar{s}+\lambda(\bar{s})t}}{\sqrt{2\pi r(\bar{s},t)}}.
\end{equation}
Next we require the solutions to reach a maximum. The function $e^{-\ln \bar{s}+y\cdot \bar{s}+\lambda(\bar{s})t}$ reaches its maximum at depth $t_{max}$ given by
\begin{equation}\label{145}
\left[-\frac{1}{\bar{s}}+y+\lambda'(\bar{s})t_{max}\right]\left(\frac{\mbox{d}\bar{s}}{\mbox{d}t}\right)_{t=t_{max}}+\lambda(\bar{s})=0.
\end{equation}
Because of the saddlepoint relation (\ref{139}) the latter implies $\lambda(\bar{s})=0$. From (\ref{138}) it is inferred that the maximum occurs at $\bar{s}=1$. It follows from (\ref{138}), (\ref{139}) and (\ref{141}) that 
\begin{equation}\label{146}
r(\bar{s},t) \approx y\sqrt{\frac{y}{t}},
\end{equation}
\begin{equation}\label{147}
\lambda(\bar{s})t \approx 2\sqrt{yt}-2t
\end{equation}
and
\begin{equation}\label{149}
\bar{s} \approx 2 \sqrt{\frac{t}{y}}-1.
\end{equation}
Hence, 
\begin{equation}\label{150}
N(W>E,t)= \sqrt{\sqrt{\frac{y}{t}}}\cdot \frac{e^{4\sqrt{ty}-2t}}{e^y 2\sqrt{2\pi y }}.
\end{equation}
First we will consider the situation without the factor $\sqrt{\sqrt{\frac{y}{t}}}$. At the end of this section we will show that  the influence of this factor can be neglected. 

For $E=E_c$ and thus $y=y_c \equiv n_c \ln 2$ we obtain for the age parameter, leaving the bar,
\begin{equation}\label{151}
s=2\sqrt{\frac{t}{n_c \ln2}}-1
\end{equation}
and for the shower size
\begin{equation}\label{152}
N(t)=N(W>E_c,t)= \frac{e^{4\sqrt{tn_c \ln 2}-2t}}{2\cdot2^{n_c}\sqrt{2\pi \cdot n_c \ln 2 }}.
\end{equation}
For the charged part this is
\begin{equation}\label{153}
N_{e\pm}(t)= \frac{e^{4\sqrt{tn_c \ln 2}-2t}}{3\cdot2^{n_c}\sqrt{2\pi \cdot n_c \ln 2 }}.
\end{equation}
The latter can also be written as
\begin{equation}\label{153a}
N_{e\pm}(t)= \frac{0.16 \cdot 2^{n_c}}{\sqrt{n_c}}e^{-2(t-2\sqrt{tn_c \ln 2}+n_c \ln 2)}.
\end{equation}
By means of the age parameter (\ref{151}) it can also be written as 
\begin{equation}\label{153b}
N_{e\pm}(t)= \frac{0.16 \cdot 2^{n_c}}{\sqrt{n_c}}e^{-\frac{1}{2} n_c \ln 2 (s-1)^2}.
\end{equation}
We clearly recognize it as a Gaussian in Age. Notice that for the present age parameter $s=-1$ if $t=0$.
Although the corresponding profile does have the right shape with a maximum at the right position, $t_{\max}=n_c\ln2$, it differs in height and width from the Greisen profile. To obtain the same height we simply replace 0.16 by 0.37. To obtain the same width we replace the factor $\frac{1}{2}$ in the exponent by $\frac{1}{3}$. Then the eqs. (\ref{153a}) and (\ref{153b}) read
\begin{equation}\label{153c}
N_{e\pm}(t)= \frac{0.37 \cdot 2^{n_c}}{\sqrt{n_c}}e^{-\frac{4}{3}(t-2\sqrt{tn_c \ln 2}+n_c \ln 2)}
\end{equation}
and 
\begin{equation}\label{153d}
N_{e\pm}(t)= \frac{0.37 \cdot 2^{n_c}}{\sqrt{n_c}}e^{-\frac{1}{3} n_c \ln 2 (s-1)^2}
\end{equation}
respectively. From equation (\ref{45}) it is inferred that the numerical value $0.37$ in the expression for the shower size is almost equal to the semi-theoretical value: $\frac{2}{3\sqrt{\pi}}$. So, we can also write the Gaussian in Age profile as 
\begin{equation}\label{153e}
N_{e\pm}(t)=\frac{2}{3 \sqrt{\pi}} \frac{2^{n_c}}{\sqrt{n_c}}e^{-\frac{1}{3} n_c \ln 2 (s-1)^2}.
\end{equation}
The latter Gaussian in Age profile has standard deviation $\sigma =\frac{\sqrt{3}}{\sqrt{2n_c \ln2}}$, by means of which the Gaussian in Age profile can also be written as
\begin{equation}\label{153f}
N_{e\pm}(t)=\frac{2}{\sqrt{3\ln 2}}\cdot  \frac{2^{n_c}}{n_c} \cdot \frac{1}{\sigma \sqrt{2\pi}} \cdot e^{-\frac{1}{2}\left(\frac{s-1}{\sigma}\right)^2}.
\end{equation}

In figure 3 both the present Gaussian in Age profile and the Greisen profile are plotted. We see the profiles match nicely.
\begin{figure}[htbp]
\includegraphics{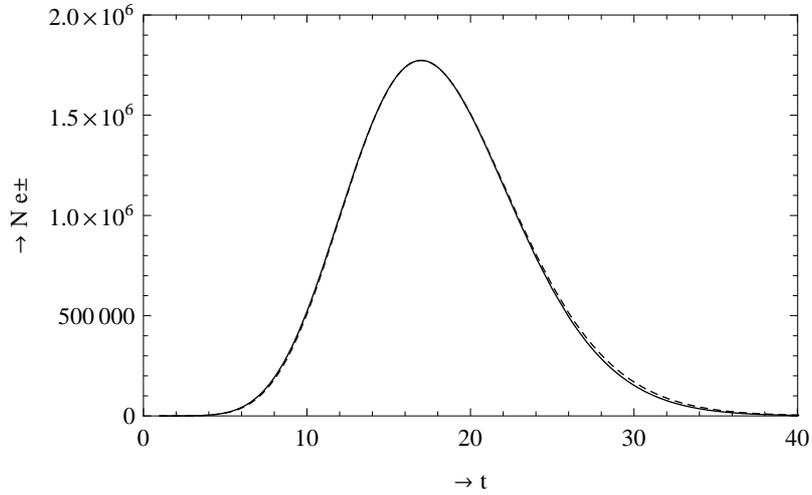}
\caption{Longitudinal shower profiles according to the present 'Gaussian in Age' function (solid) and the Greisen function (dashed) for a $2 \cdot 10^{15}$ eV shower. Vertical is the number of electrons and positrons and horizontal the atmospheric depth (in units of radiation length).}
\end{figure}
For practical purposes the Gaussian in Age profile can be generalized to a three parameter trial function
\begin{equation}\label{153g}
N_{e\pm}(t)= N_{max}\cdot e^{-w \left( \sqrt{t}-\sqrt{t_{max}} \right)^2}.
\end{equation}
The parameter $w$ determines the width of the profile; its value will be close to $\frac{4}{3}$. As for the Gaiser-Hillas function, it can be generalized further to a four parameter function by means of the shift $t \rightarrow t-t_0$ and $t_{max} \rightarrow t_{max}-t_0$.
Finally we will consider the situation where the factor $\sqrt{\sqrt{\frac{y}{t}}}$ is not neglected. Then the expression (\ref{153c}) should be modified accordingly. That is, (\ref{153c}) should be muliplied by a factor $\left(\frac{n_c \ln 2}{t} \right)^\frac{1}{4}$:
\begin{equation}\label{153h}
N_{e\pm}(t)= \frac{0.37 \cdot 2^{n_c}}{\sqrt{n_c}}\left(\frac{n_c \ln 2}{t} \right)^\frac{1}{4} \cdot e^{-\frac{4}{3}(t-2\sqrt{tn_c \ln 2}+n_c \ln 2)}
\end{equation}
In figure 4 both the profile according to (\ref{153h}) and the Greisen profile are plotted.
\begin{figure}[htbp]
\includegraphics{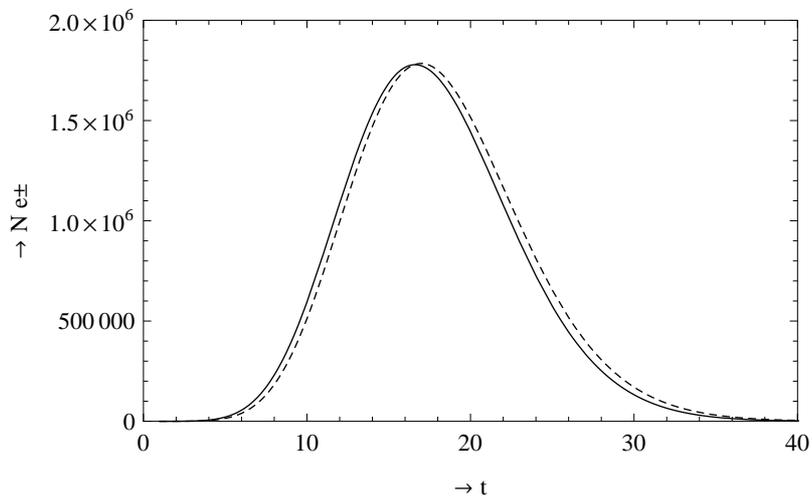}
\caption{Longitudinal shower profiles according to the expression (\ref{153h}) (solid) and the Greisen function (dashed) for a $2 \cdot 10^{15}$ eV shower. Vertical is the number of electrons and positrons and horizontal the atmospheric depth (in units of radiation length).}
\end{figure}
The influence of the factor is small and practically equal to a shift over a distance $\frac{1}{2}d$. As mentioned before, a small translation is not of interest for the present comparison of profiles and will therefore be ignored. 

Although redundant a similar analysis as in this section can in principle be applied to the eq. (\ref{63}) as well. Such an analysis leads exactly to the solution (\ref{18}) for the number of particles in the discrete energy levels and approximately to the Gamma distribution (\ref{27h}). In fact this was our main motivation to take the Gamma distribution for the total number of particles in section 4. 

\section{Summary and conclusions}
To day both the Gaisser-Hillas function and the Gaussian in Age function are used as trial functions for the reconstruction of longitdinal shower profiles \cite{{11},{12}}. There are even trial functions composed of two halves of Gaussian in Age functions \cite{13}. All the these function usually contain parameters which are not independent of each other \cite{14}. Recently new less dependent parameters for the Gaisser-Hillas and the Gaussian in Age functions were constructed \cite{{14},{15}}. 
By introducing the parameter $\mu =y_c \equiv n_c \ln 2$, the expressions for the shower profile the espression for the shower profile can be written by means of a single parameter. The Greisen function, the Gaisser-Hillas function and the Gaussian in Age function then respectively read 
\begin{equation}\label{154}
N_{e\pm}(t)= \frac{0.31}{\sqrt{\mu}}\cdot \left( \frac{1}{3}+\frac{2}{3} \frac{\mu}{t}\right)^{\frac{3}{2}t}\cdot e^{t},
\end{equation}

\begin{equation}\label{155}
N_{e\pm}(t)= \frac{0.31}{\sqrt{\mu}}\cdot \left( \frac{t}{\mu}\right)^{\frac{2}{3}\mu}\cdot e^{\frac{5}{3}\mu-\frac{2}{3}t},
\end{equation}
and 
\begin{equation}\label{156}
N_{e\pm}(t)= \frac{0.31}{\sqrt{\mu}}\cdot e^{-\frac{1}{3}\mu+\frac{8}{3}\sqrt{\mu t}-\frac{4}{3}t}.
\end{equation}
In this minimal form all three profiles practically coincide and reach a maximum $\frac{0.31}{\sqrt{\mu}}\cdot e^{\mu}$ at depth $t_{max}=\mu$. If desired one can replace the numerical constant 0.31 by its semi-theoretical analogon $\frac{2 \sqrt{ \ln 2}}{3\sqrt{\pi}}$.
It is readily admitted that we crudely neglected small translations and ignored the differences between individual shower profiles in order to obtain the single parameter expressions. Of course, to take account for small translations and other details of observed or simulated individual shower profiles additional parameters are unavoidable. 

The fact that the three profile functions practically coincide gives rise to the idea that there must be a mathematical connection or common origin. We found the connection by solving the Rossi and Greisen equations for simplified cross sections. The similarity of the three profiles also leads to the conclusion that the shape of the shower profile is rather independent of the type of cross section. Instead we conclude that the characteristic shape of the shower profile is governed by the statistics of the splittings. Only splittings with substantial different probability distributions, such as in the Heitler model, will lead to a substantial different profile. The functional form of the cross sections mainly influence the height and the width of the profile.

Because of its mathematical convenience the Gaiser-Hillas function is sometimes favoured over the Greisen function \cite{16}. The Gaussian in Age function is mathematically convenient as well. The Greisen function is less convenient. The points of inflection, for instance, are for the  Gaiser-Hillas function the roots $\mu\pm \sqrt{\frac{3\mu}{2}}$ of the quadratic equation $t^2-2\mu t+\mu^2-\frac{3}{2}\mu=0$. For the Gaussian in Age function the points of inflection are two of the roots of the cubic equation $t^3-2\mu t^2+(\mu^2-\frac{3}{2}\mu)t-\frac{9}{64}\mu=0$. The degree of these polynomials expresses the hierarchy in the complexity of the corresponding expressions for the shower profiles. For the Greisen function the points of inflection can not be derived analytically.

In trial functions for longitudinal shower profiles the Greisen age parameter 
\begin{equation}\label{158}
s=\frac{3t}{t+2t_{max}}
\end{equation}
is commonly used. However, one should be careful with regarding it as a universal age parameter since the functional form of the age parameter is model dependent \cite{17}. With the use of the parameter $\mu$ the Gaussian in Age function reads
\begin{equation}\label{159}
N_{e\pm}(t)= \frac{0.31\cdot e^{\mu}}{\sqrt{\mu}}\cdot e^{-\frac{1}{3}\mu(s-1)^2},
\end{equation}
while the present analysis suggests
\begin{equation}\label{160}
s=2\sqrt{\frac{t}{t_{max}}}-1
\end{equation}
as the natural age parameter.
It seems worthwhile to investigate the alternative age parameter on its practical use. This subject is currently under research.
\\

\newpage

{\Large \textbf{Acknowledgements}}\\
I am grateful to the reviewers for their valuable suggestions and improvements. I wish to thank Prof. J.W. van Holten and Prof. B. van Eijk for their encouragement and support. I also wish to thank Nikhef for its hospitality. The work is supported by a grant from FOM (Foundation for Fundamental Research on Matter).

\bibliography{References}

\end{document}